\documentclass{elsart}

\usepackage{amsmath}
\usepackage{amssymb}

\begin{document}
\begin{frontmatter}
\begin{flushleft}
TTP08-19\\
SFB/CPP-08-29\\
arXiv:0806.3405
\end{flushleft}
\title{The second physical moment of the heavy quark vector correlator at $\mathcal{O}(\alpha_s^3)$} 
\author[Karlsruhe]{A.~Maier}, 
\author[Karlsruhe]{P.~Maierh\"ofer} and 
\author[Karlsruhe]{P.~Marquard}
\address[Karlsruhe]{Institut f\"ur Theoretische Teilchenphysik,
Universit\"at Karlsruhe, 76128 Karlsruhe, Germany}
\begin{abstract}
The second moment of the heavy quark vector correlator at ${\cal O}(\alpha_s^3)$ is presented. The implications of this result on recent determinations of the charm and bottom quark mass are discussed.
\end{abstract}
\begin{keyword}
Perturbative calculations, Quantum
Chromodynamics,
Dispersion Relations, Charm Quarks, 
Bottom Quarks
\PACS 12.38.Bx, 12.38.-t,
11.55.Fv, 14.65.Dw, 14.65.Fy
\end{keyword}
\end{frontmatter}

\section{Introduction}\label{sec:introduction}

Correlators of quark currents are of prime interest for several
phenomenological applications. Their low-energy expansions, in
particular, allow for the precise determination of charm and bottom
quark masses via QCD sum rules
\cite{Shifman:1978by,Reinders:1984sr,Kuhn:2001dm,Kuhn:2007vp,Allison:2008xk}.
For this reason, heavy quark correlators have been frequently
investigated in the framework of perturbation theory.

Up to ${\cal O}(\alpha_s^2)$, analytic expansions to great depth are
known for the low energy region. The three-loop QCD corrections to the
correlator of two vector currents were first calculated in
\cite{Chetyrkin:1995ii}. In \cite{Chetyrkin:1997mb} up to seven terms in
the low energy expansion were obtained. This calculation also included
further currents, namely the scalar, pseudo-scalar, and axial-vector
current. Recently the calculation at three-loops has been extended to
moments up to $n=30$ for all four currents
\cite{Boughezal:2006uu,Maier:2007yn}.

The moments of the vector correlator can then be used to extract the
value of the masses of the charm and bottom quark from $e^+e^-$ data in
the threshold region using the $R$-ratio, since they are related via a
dispersion relation. A brief outline of this method is given in Section
\ref{sec:notation}, which was first applied at three loops in
\cite{Kuhn:2001dm}.

At three loops a significant, sometimes dominant part of the error
arises from the theoretical uncertainty due to higher orders, often
estimated by the renormalization scale dependence.  Therefore the
calculation had to be taken to the four-loop level
\cite{Chetyrkin:2006xg,Boughezal:2006px} to reach a precision comparable
to or below the experimental data. The contributions from
double-fermionic loop insertions of heavy and/or light quarks are known
explicitely up to 30 terms in the low energy expansion
\cite{Czakon:2007qi}. The contributions due to light quark loop
insertions of ${\cal O}(\alpha_s^n n_l^{n-1})$ are known to all orders
in $\alpha_s$ \cite{Grozin:2004ez}. Recently the lower moments were also
calculated for the remaining three currents in \cite{Sturm:2008eb}.

In \cite{Kuhn:2007vp} the first moment of the vector correlator was used
to extract the masses of the charm and bottom quarks. Since all but constant
terms are known from renormalization group arguments, the analysis was
done for up to the fourth moment, employing a conservative error
estimate for the missing constant terms.

In this paper we present the calculation of the second moment of the
vector correlator and discuss its impact on the determination of the
charm and bottom quark masses.


The outline of this paper is as follows: In Section
\ref{sec:notation} we set the framework and notations used throughout
the paper. In Section~\ref{sec:calculation-results} we explain
the details of the calculation, present the result for the second physical
moment and discuss its impact on the the quark mass determination. A
brief summary and conclusions are given in Section \ref{sec:conclusion}.

\section{Notation}\label{sec:notation}

The correlator $\Pi^{\mu\nu}(q)$ of two vector currents is defined as
\begin{equation}
  \label{eq:1}
\Pi^{\mu\nu}(q) = i \int dx \, e^{iqx} \langle 0|Tj^\mu(x) j^\nu(0) |0
\rangle \, ,
\end{equation}
with the current $j^\mu(x) = \bar \Psi(x) \gamma^\mu \Psi(x)$ being
composed of the heavy quark fields $\Psi(x)$. The function
$\Pi^{\mu\nu}(q)$ is conveniently written in the form
\begin{equation}
  \label{eq:2}
  \Pi^{\mu\nu}(q) = (-q^2 g^{\mu\nu} + q^\mu q^\nu ) \Pi(q^2) \, .
\end{equation}
It can be related to the ratio $R(s) = \sigma(e^+ e^-
\to \mbox{hadrons})/\sigma(e^+ e^- \to \mu^+ \mu^-)$ with the help of
the dispersion relation 
\begin{equation}
  \label{eq:3}
  \Pi(q^2) = \frac{1}{12 \pi^2} \int _0 ^\infty ds \frac{R(s)}{s\,(s-q^2)} \, ,
\end{equation}
where the normalization $\Pi(0)=0$ has been adopted. 

To extract the quark masses the experimental data on the right hand side
of (\ref{eq:3}) has to be compared with the theoretical evaluation of
$\Pi(q^2)$ on the left hand side. This is best be done by comparing the
corresponding Taylor series in $q^2$.  The $n$-th derivatives with respect to $q^2$
at $q^2=0$ define the experimental moments
\begin{equation}
  \label{eq:4}
  {\cal M}_n^{exp} = \int ds \frac{R(s)}{s^{n+1}} \, ,
\end{equation}
which can be compared with the theoretical moments
\begin{equation}
  \label{eq:5}
  {\cal M}_n^{th} = Q_q^2 \frac{9}{4} \left ( \frac{1}{4 \bar m_q^2}
  \right )^n \bar C_n \, .
\end{equation}
The latter are related to the Taylor coefficients $\bar C_n$ of the vacuum
polarization function
\begin{equation}
  \label{eq:6}
  \bar \Pi (q^2) = \frac{3 Q_q^2}{16 \pi^2} \sum_{n\geq 0} \bar C_n \bar
  z^n 
\end{equation}
with $\bar z = q^2 / (4 \bar m^2)$. Symbols carrying a bar indicate that the renormalization has been performed in the $\overline{\mbox{MS}}$ scheme. The coefficients $\bar{C}_n$ can be expanded in a power series in
$\frac{\alpha_s}{\pi}$
\[
\bar{C_n} = \bar C_n^{(0)} + \frac{\alpha_s}{\pi} \bar C_n^{(1)} + \left(
  \frac{\alpha_s}{\pi}\right ) ^2 \bar C_n^{(2)} + \left(
  \frac{\alpha_s}{\pi}\right ) ^3 \bar C_n^{(3)} + \cdots \, .
\]
The four-loop contribution $\bar C_n^{(3)}$ can be decomposed according
to the number of quark loops and colour structures as follows:
\begin{align}
\label{eq:9} 
\bar{C}_n^{(3)} = &C_F T_F^2 n_l^2 \bar{C}_{ll,n}^{(3)} 
+ C_F T_F^2 n_h^2 \bar{C}_{hh,n}^{(3)}
+ C_F T_F^2 n_l n_h \bar{C}_{lh,n}^{(3)} \nonumber \\
 &+  C_F T_F n_l\left ( C_A \bar{C}_{lNA,n}^{(3)} + C_F
   \bar{C}_{lA,n}^{(3)} \right ) +  \bar{C}_{n_f^0,n}^{(3)} \\
 &+  C_F T_F n_h\left( C_A \bar{C}_{hNA,n}^{(3)} + C_F
   \bar{C}_{hA,n}^{(3)} \right )+
 \frac{n_h}{N_C}d^{abc}d^{abc}\bar{C}_{S,n}^{(3)}\nonumber \, .
\end{align}
$\bar{C}_{n_f^0,n}^{(3)}$ contains the purely bosonic contributions,
where we set the number of colours $N_C=3$ for simplicity, while
$\bar{C}_{S,n}^{(3)}$ denotes the contribution from singlet diagrams.
$C_F=\frac{N_C^2 - 1}{2N_C}$ and $C_A = N_C$ are the Casimir operators
of the fundamental and adjoint representation of the $SU(N_C)$ group,
respectively. $T_F=\frac{1}{2}$ is the index of the fundamental representation.
$d^{abc}$ is the symmetric structure constant. $n_l$ and
$n_h=1$ denote the number of light and heavy quarks, respectively.

\section{Calculation and Results}\label{sec:calculation-results}

The diagrams have been generated using {\tt QGRAF}
\cite{Nogueira:1991ex}. Expanding them in $q^2$ results in four-loop
tadpole integrals. Using {\tt EXP} \cite{exp} they are mapped to six
topologies with the maximum of nine lines. The main difficulty of  the
calculation lies in the reduction of the vast amount of integrals to
 the small set of 13 master integrals. This is done using  Integration-By-Parts
identities \cite{Chetyrkin:1981qh} together with  the Laporta
algorithm \cite{Laporta:2001dd} which is efficiently  implemented in the
multi-threaded C++
 program {\tt CRUSHER} \cite{crusher}. {\tt CRUSHER} uses \texttt{GiNAC} \cite{Bauer:2000cp}
 for simple algebraic manipulations and \texttt{Fermat} \cite{fug} for the
 simplification of complicated ratios of polynomials. 
A supplementary
technique to perform the reduction to master integrals is based on the
idea that self energy subgraphs of the integral can be reduced
independently in order to effectively reduce the number of loops of the
diagram. This can be useful because these integrals have up to two more
propagator powers than integrals without an internal self energy and are
therefore more cumbersome for traditional Laporta algorithm. In
combination with Groebner Bases and the \texttt{Mathematica} package
\texttt{FIRE} \cite{Smirnov:2005ky,Smirnov:2006tz,Smirnov:fire} it is
also possible to calculate integrals without internal self energies.
A more detailed description of the calculation techniques will be published
soon \cite{sepaper}. In total the reduction of 1.8 million integrals was
needed in order to perform the calculation, which  is done 
using {\tt FORM} \cite{Vermaseren:2000nd} in combination with
the MATAD \cite{Steinhauser:2000ry} setup. The necessary master
integrals have been calculated in
\cite{Laporta:2002pg,Chetyrkin:2004fq,Kniehl:2005yc,Schroder:2005db,Bejdakic:2006vg,Kniehl:2006bf,Kniehl:2006bg}.
We confirm the results for the zeroth and first moment given in
\cite{Boughezal:2006uu,Chetyrkin:2006xg,Boughezal:2006px}.

Inserting the master integrals and performing the renormalization of the strong coupling constant and the mass in the $\overline{\mbox{MS}}$ scheme leads to the following result for the second moment at $\mu^2 = m^2$ as defined in Eq. (\ref{eq:9}):
{\small
\begin{align*}
\bar{C}^{(3)}_{n_f^0,2}=&
+\frac{64985074258811347}{353072079360000}-\frac{2900811008}{3648645}a_5
\\&-\frac{1662518706713}{21016195200}\left(24 a_4+\log^4 2-6\zeta_2\log^2 2\right)
 +\frac{362601376}{54729675}\log^5 2\\&-\frac{725202752}{10945935}\zeta_2\log^3 2-\frac{1684950406}{3648645}\zeta_4\log2\\
 & +\frac{112680551036302633}{47076277248000}\zeta_3-\frac{26401638588211}{28021593600}\zeta_4-\frac{164928917}{270270}\zeta_5\;,\displaybreak[0]\\
\bar{C}^{(3)}_{S,2} =
&+\frac{5881974201847}{8369115955200}+\frac{97011619}{696729600}\left(24
  a_4+\log^4 2-6\zeta_2\log^2 2\right)\\ &+\frac{796232393699}{371960709120}\zeta_3
 -\frac{745372259}{185794560}\zeta_4\;,\displaybreak[0]\\
\bar{C}^{(3)}_{hNA,2} =
&-\frac{20427854209619}{5649153269760}-\frac{31595849}{11612160}\left(24
  a_4+\log^4 2-6\zeta_2\log^2 2\right)\\& -\frac{29638030087837}{697426329600}\zeta_3
 +\frac{968787977}{15482880}\zeta_4+\frac{362}{63}\zeta_5\;,\displaybreak[0]\\
\bar{C}^{(3)}_{lNA,2}= &-\frac{22559166733}{16796160000}-\frac{520999}{4354560}\left(24 a_4+\log^4 2-6\zeta_2\log^2 2\right)\\&-\frac{309132631}{12902400}\zeta_3+\frac{167529079}{5806080}\zeta_4\;,\displaybreak[0]\\
\bar{C}^{(3)}_{hA,2}= &-\frac{37320009196157}{271593907200}-\frac{130387543}{2177280}\left(24 a_4+\log^4 2-6\zeta_2\log^2 2\right)\\&-\frac{5811074101069}{6706022400}\zeta_3
 +\frac{2218910663}{1451520}\zeta_4\;,\displaybreak[0]\\
\bar{C}^{(3)}_{lA,2}= &+\frac{357543003871}{11757312000}+\frac{520999}{2177280}\left(24 a_4+\log^4 2-6\zeta_2\log^2 2\right)\\&-\frac{36896356307}{174182400}\zeta_3+\frac{598455689}{2903040}\zeta_4\;,\displaybreak[0]\\
\bar{C}^{(3)}_{lh,2}= &+\frac{95040709}{62705664}-\frac{2029}{41472}\left(24 a_4+\log^4 2-6\zeta_2\log^2 2\right)\\&-\frac{12159109}{4644864}\zeta_3+\frac{99421}{55296}\zeta_4\;,\displaybreak[0]\\
\bar{C}^{(3)}_{hh,2}= &+\frac{1842464707}{646652160}-\frac{2744471}{1064448}\zeta_3\;,\displaybreak[0]\\
\bar{C}^{(3)}_{ll,2}=
&+\frac{15441973}{19136250}-\frac{32}{45}\zeta_3 \, ,
\end{align*}}~\\[-2ex]
where Riemann's zeta function $\zeta_n$ and the polylogarithm $\mathrm{Li}
_n(1/2)$ are defined by
\begin{equation}
  \label{eq:7}
  \zeta_n = \sum_{k=1}^\infty \frac{1}{k^n} \quad \mbox{and} \quad
  a_n=\mathrm{Li}_n(1/2) = \sum_{k=1} ^ \infty \frac{1}{2^k k^n} \, . 
\end{equation}
For completeness we also give the results for the singlet contribution
to the zeroth and first moment:
\begin{align}
\bar{C}^{(3)}_{S,0} =
&\frac{2411}{20160}-\frac{6779}{4480}\zeta_3+\frac{2189}{768}\zeta_4-\frac{5}{48}\zeta_5-\frac{73}{576}\left(24
  a_4+\log^4 2-6 \zeta_2 \log^2 2\right)\, ,\\
\bar{C}^{(3)}_{S,1} = &
\frac{664837}{2566080}-\frac{2017831}{855360}\zeta_3+\frac{175}{48}\zeta_4-\frac{739}{4320}\left(24
  a_4+\log^4 2-6 \zeta_2 \log^2 2\right) \, .
\end{align}
Numerically at $\mu^2=m^2$ one finds $\bar C_2^{(3)} |_{n_l=3} = -3.49373 + 0.155877$ and $\bar C_2^{(3)} |_{n_l=4} = -2.64381 + 0.155877$. The second term in each of these equations corresponds to the singlet contribution.

Extracting the charm and bottom quark mass from the second moment using the input data given in \cite{Kuhn:2007vp} with the new value of $C_2^{(3)}$ leads to a shift of $-3\,$MeV for $m_c$ and $-2\,$MeV for $m_b$ and yields
\begin{align}
m_c(3\, \mathrm{ GeV}) = 0.976(16)\, \mathrm{ GeV} \quad\mbox{and}\quad m_b(10\, \mathrm{ GeV}) = 3.607(19)\, \mathrm {GeV} \ .
\end{align}
This can be converted to the values at $m_c$ and $m_b$, $m_c(m_c) = 1.277(16)\,\mathrm{GeV}$ and $m_b(m_b) = 4.162(19)\,\mathrm{GeV}$, respectively.

The final results for the quark masses given in \cite{Kuhn:2007vp} are
$m_b(m_b)=4.164(25)\,\mathrm{GeV}$ and
$m_c(m_c)=1.286(13)\,\mathrm{GeV}$, respectively. In case of $m_c$ the
first moment was used at $\mathcal{O}(\alpha_s^3)$ accuracy. For $m_b$
the second moment, which was known only up to $\mathcal{O}(\alpha_s^2)$
at that time, was chosen. In the latter case the logarithms at
$\mathcal{O}(\alpha_s^3)$ calculated by means of renormalization group
methods were included and the error estimate was based on the missing
constant term. Although this estimate was based on plausible arguments
only a real calculation could prove its validity.
Removing the $6$~MeV error, which arises from the estimated term in case
of the $b$ quark, the total error of $m_b$ is reduced by $\sim{}25\,\%$.
In order $\alpha_s^3$ the perturbative error is practically negligible
and the remaining $19$~MeV error arises from the experimental
uncertainty and from the value of $\alpha_s$. At present this is the
most precise determination of the bottom quark mass.

As already discussed in \cite{Kuhn:2007vp}, different moments weight the
experimental results from larger and smaller $s$ values differently.
Therefore it is important to compare the obtained quark masses from
several moments to test the self-consistency of the method and the
stability of the results. Because of sparse and poor experimental
data in the continuum region above $4.8\,\mbox{GeV}$ (for $m_c$) and
$11.2\,\mbox{GeV}$ (for $m_b$), the data for $R(s)$ were replaced by
perturbative QCD in the analysis. This region can be suppressed by using
higher moments, which is especially important in the case of $m_b$ where
the first moment, which was already under full theoretical control at
order $\alpha_s^3$ in \cite{Kuhn:2007vp}, receives a large contribution
from the region above $11.2\,\mbox{GeV}$. The situation is significantly
better for the second moment, which is now also fully under control from
the theory side. For the determination of $m_c$ the first and the second
moment are of equal reliability and the consistency between the two
results for $m_c(3\,\mbox{GeV})$, namely $0.986(13)\,\mbox{GeV}$ and
$0.976(16)\,\mbox{GeV}$, is remarkable. On the other hand for higher
moments non-perturbative effects increase (especially for $m_c$) leading
to larger theoretical uncertainties. For these reasons we think that for
$m_b$ the second or maybe third moment are best suited for the mass
determination, while for $m_c$ the first and second moment are
preferred.

Apart from the application discussed above, the higher moments evaluated
above have been used recently for quark mass determinations from lattice
simulations \cite{Allison:2008xk} and for the reconstruction of the full
$q^2$ dependence of the vacuum polarization at $\mathcal{O}(\alpha_s^3)$
\cite{Hoang:2008qy}.

\section{Summary and Conclusion}\label{sec:conclusion}

We have presented the second physical moment in the low energy expansion
of the heavy quark vector correlator at four-loop order, including the
singlet contribution. Although this contribution only causes a rather
small shift in the quark masses obtained from the second moment the
error is reduced significantly. The values remain in good agreement with
those extracted using the first moment.

\section*{Acknowledgements}\label{sec:acknowledgements}

We thank K.\,G.~Chetyrkin, J.\,H.~K\"uhn and M.~Steinhauser for helpful discussions and cross checks. We also like to thank A.\,V.~Smirnov for providing us with the \texttt{FIRE} package and extending it for our purposes and for interesting discussions.

This work was supported by the Deutsche Forschungsgemeinschaft through the SFB/TR-9 ``Computational Particle Physics''. A.\,M. and Ph.\,M. were supported by the Graduiertenkolleg ``Hochenergiephysik und Teilchenastrophysik''. A.\,M. thanks the Landesgraduiertenf\"orderung for support.


\end{document}